\documentclass{article}
\usepackage{frascatiphys}
\usepackage{graphicx}
\usepackage{dcolumn}% Align table columns on decimal point
\usepackage{bm}% bold math
\usepackage{amsmath}
\usepackage{amssymb}
\usepackage{url}
\usepackage[normalem]{ulem}
\usepackage{xcolor}
\usepackage{listings} 
\begin{document}
\title{ 
$\rho$-MESON LEPTOPRODUCTION AS TESTFIELD FOR THE UNINTEGRATED GLUON DISTRIBUTION IN THE PROTON
}
\author{
Andr\`ee Dafne Bolognino          \\
{\em Dipartimento di Fisica dell'Universit\`a della Calabria
	I-87036 Arcavacata di Rende, Cosenza, Italy} \\
{\em INFN - Gruppo collegato di Cosenza, I-87036 Arcavacata di Rende,
	Cosenza, Italy}\\
Francesco Giovanni Celiberto       \\
{\em Instituto de Fi\'sica Te\'orica UAM/CSIC, Nicol\'as Cabrera 15, 28049 Madrid, Spain}\\
{\em Universidad Aut\'onoma de Madrid, 28049 Madrid, Spain}\\
Dmitry Yu. Ivanov                  \\
{\em Sobolev Institute of Mathematics, 630090 Novosibirsk, Russia}\\
{\em Novosibirsk State University, 630090 Novosibirsk, Russia}\\
Alessandro Papa                    \\
{\em Dipartimento di Fisica dell'Universit\`a della Calabria I-87036 Arcavacata di Rende, Cosenza, Italy} \\
{\em INFN - Gruppo collegato di Cosenza, I-87036 Arcavacata di Rende,
	Cosenza, Italy}
}
\maketitle
\baselineskip=10pt
\begin{abstract}
The gluon content of the proton is embodied by the unintegrated gluon distribution (UGD), which has universal validity. In literature many models of UGD have been proposed so far.
The polarized $\rho$-meson leptoproduction at HERA offers a nowadays unexplored testfield to discriminate among existing models of UGD, via the comparison with theoretical predictions formulated in the $\kappa$-factorization approach.
\end{abstract}
\baselineskip=14pt

\section{Introduction}
\label{introd}

Our ability to find new Physics at the LHC strongly relies on getting a
more and more precise understanding on the structure of the proton.
In approaches based on the theoretical study of collision processes, the
information about the proton structure is encoded in the partonic distribution
functions entering the \emph{factorized} expression for the cross section. At small $x$ the proper factorization scheme
is given by {\em $\kappa$-factorization}, whereby the DIS cross section is
represented as the convolution of the unintegrated gluon distribution (UGD)
in the proton with the (perturbative) impact factor (IF) for the $\gamma^*
\to \gamma^*$ transition. The UGD is a nonperturbative quantity, function of
$x$ and $\kappa$, where the latter represents the gluon momentum transverse
to the direction of the proton. The UGD, in its original definition, obeys the
BFKL~\cite{BFKL} evolution equation in the $x$ variable.
The UGD is not well known and several
models for it, which lead to very different shapes in the
$(x,\kappa)$-plane, have been proposed so far (see, for instance, Ref.~\cite{small_x_WG}).

The aim of this work is to present some arguments that HERA data on
polarization observables in vector meson (VM) electroproduction can be
used to constrain the $\kappa$ dependence of the UGD in the HERA energy range.
In particular, we will focus our attention on the {\em ratio} of the two
dominant amplitudes for the polarized electroproduction of $\rho$ mesons,
{\it i.e.} the longitudinal VM production from longitudinally polarized
virtual photons and the transverse VM production from transversely polarized
virtual photons. 

The paper is organized as follows: in Section~\ref{theory} we will present
the expressions for the amplitudes of interest here, and sketch the main properties of
a few models for the UGD; in Section~\ref{analysis} we compare theoretical
predictions from the different models of UGD with HERA data; in
Section~\ref{discussion} we draw our conclusions.

\section{Theoretical setup}
\label{theory}

The H1 and ZEUS collaborations have provided useful and complete analyses
of the hard exclusive production of the $\rho$ meson in $ep$ collisions
through the subprocess
\begin{equation}
	\label{process}
	\gamma^*(\lambda_\gamma)p\rightarrow \rho (\lambda_\rho)p\,.
\end{equation}
Here $\lambda_\rho$ and $\lambda_\gamma$ represent the meson and photon
helicities, respectively, and can take the values 0 (longitudinal polarization)
and $\pm 1$ (transverse polarizations).  The helicity amplitudes 
$T_{\lambda_\rho \lambda_\gamma}$ extracted at HERA~\cite{Aaron:2009xp} exhibit the
following hierarchy~\cite{Ivanov:1998gk}: 
\begin{equation}
	T_{00} \gg T_{11} \gg T_{10} \gg T_{01} \gg T_{-11}.
\end{equation}
The H1 and ZEUS collaborations have analyzed data in different ranges of $Q^2$
and $W$. In what follows we will refer only to the H1 ranges,
\begin{equation}
	\begin{split}
		2.5\,\text{\rm GeV$^2$} < Q^2 <60 \,\text{\rm GeV$^2$},\\
		35\, \text{GeV} < W < 180\,\text{GeV},
	\end{split}
\end{equation}
and will concentrate only on the dominant helicity amplitude ratio, $T_{11}/T_{00}$.

\subsection{Electroproduction of polarized $\rho$ mesons in the
	$\kappa$-factorization}
\label{leptoproduction}

In the high-energy regime, $s\equiv W^2\gg Q^2\gg\Lambda_{\rm QCD}^2$, which
implies small $x=Q^2/W^2$, the forward helicity amplitude for the
$\rho$-meson electroproduction can be written, in $\kappa$-factorization, as
the convolution of the $\gamma^*\rightarrow \rho$ IF,
$\Phi^{\gamma^*(\lambda_\gamma)\rightarrow\rho(\lambda_\rho)}(\kappa^2,Q^2)$,
with the UGD, ${\cal F}(x,\kappa^2)$. Its expression reads
\begin{equation}
	\label{amplitude}
	T_{\lambda_\rho\lambda_\gamma}(s,Q^2) = \frac{is}{(2\pi)^2}\int \dfrac{d^2\kappa}
	{(\kappa^2)^2}\Phi^{\gamma^*(\lambda_\gamma)\rightarrow\rho(\lambda_\rho)}(\kappa^2,Q^2)
	{\cal F}(x,\kappa^2),\quad \text{$x=\frac{Q^2}{s}$}\,.
\end{equation}

The expressions for the IFs, for the longitudinal and the transverse cases, take the
form given by Eq.~(33) and Eq.~(38) in Ref.~\cite{Anikin:2009bf}.
In particular, the longitudinal IF encompasses the twist-2 distribution amplitude (DA)~\cite{Ball:1998sk}; while the transverse IF is defined using DAs which encompass both genuine twist-3 and Wandzura-Wilczek (WW) contributions~\cite{Ball:1998sk, Anikin:2011sa}.
In this work we focus on the WW contributions and then we relax this approximation including the genuine parts. Moreover, for the sake of semplicity, we have adopted the \emph{asymptotic} choice for the twist-2 DA (for further details see Section 2.2 of Ref.~\cite{Bolognino:2018}).

\subsection{Models of Unintegrated Gluon Distribution}
\label{models}

A selection of six UGD models has been considered, without
pretension to exhaustive coverage, but with the aim of comparing different approaches. We refer the reader to
the original papers for details on the derivation of each model and limit
ourselves to presenting here just the functional form ${\cal F}(x,\kappa^2)$
of the UGD as we implemented it in the numerical analysis.

\subsubsection{An $x$-independent model (ABIPSW)}

The simplest UGD model is $x$-independent and merely coincides with
the proton impact factor~\cite{Anikin:2011sa}:
\begin{equation}
	{\cal F}(x,\kappa^2)= \frac{A}{(2\pi)^2\,M^2}
	\left[\frac{\kappa^2}{M^2+\kappa^2}\right]\,,
\end{equation}
where $M$ corresponds to the non-perturbative hadronic scale. The constant $A$
is unessential since we are going to consider the ratio $T_{11}/T_{00}$.

\subsubsection{Gluon momentum derivative}

This UGD is given by
\begin{equation}
	\label{xgluon}
	{\cal F}(x, \kappa^2) = \frac{dxg(x, \kappa^2)}{d\ln \kappa^2}
\end{equation}
and encompasses the collinear gluon density $g(x, \mu_F^2)$, taken at
$\mu_F^2=\kappa^2$. It is based on the obvious requirement that, when
integrated over $\kappa^2$ up to some factorization scale, the UGD must
give the standard gluon density. 
%We have employed the CT14 parametrization~\cite{Dulat:2015mca}, using the appropriate cutoff $\kappa_{\rm min} = 0.3$~GeV (see Section~III A of Ref.~UGD for further details).

\subsubsection{Ivanov--Nikolaev' (IN) UGD: a soft-hard model}

The UGD proposed in Ref.~\cite{Ivanov:2000cm} is developed with the purpose
of probing different regions of the transverse momentum. In the large-$\kappa$
region, DGLAP parametrizations for $g(x, \kappa^2)$ are employed. Moreover,
for the extrapolation of the hard gluon densities to small $\kappa^2$, an
Ansatz is made~\cite{Nikolaev:1994cd}, which describes the color gauge
invariance constraints on the radiation of soft gluons by color singlet
targets. The gluon density at small $\kappa^2$ is supplemented by a
non-perturbative soft component, according to the color-dipole
phenomenology.

This model of UGD has the following form:
\begin{equation}
	{\cal F}(x,\kappa^2)= {\cal F}^{(B)}_\text{soft}(x,\kappa^2) 
	{\kappa_{s}^2 \over 
		\kappa^2 +\kappa_{s}^2} + {\cal F}_\text{hard}(x,\kappa^2) 
	{\kappa^2 \over 
		\kappa^2 +\kappa_{h}^2}\,,
	\label{eq:4.7}
\end{equation}
 We refer the reader to Ref.~\cite{Ivanov:2000cm} for an exaustively discussion on the features, the parameters and on the expressions of the soft and the hard terms.
We stress that this model was successfully tested on the {\em unpolarized}
electroproduction of VMs at HERA.

\subsubsection{Hentschinski-Salas--Sabio Vera' (HSS) model}

This model, originally used in the study of DIS structure
functions~\cite{Hentschinski:2012kr}, takes the form of a convolution between
the BFKL gluon Green's function and a LO proton impact factor. It has been
employed in the description of single-bottom quark production at LHC
in~\cite{Chachamis:2015ona} and to investigate the photoproduction of
$J/\Psi$ and $\Upsilon$ in~\cite{Bautista:2016xnp}. We implemented
the formula given in~\cite{Chachamis:2015ona} (up to a $\kappa^2$ overall
factor needed to match our definition), which reads
\begin{equation}
	\label{HentsUGD}
	{\cal F}(x, \kappa^2, M_h) = \int_{-\infty}^{\infty}
	\frac{d\nu}{2\pi^2}\ {\cal C} \  \frac{\Gamma(\delta - i\nu -\frac{1}{2})}
	{\Gamma(\delta)}\ \left(\frac{1}{x}\right)^{\chi\left(\frac{1}{2}+i\nu\right)}
	\left(\frac{\kappa^2}{Q^2_0}\right)^{\frac{1}{2}+i\nu}
\end{equation}
\[
\times \left\{ 1 +\frac{\bar{\alpha}^2_s \beta_0 \chi_0\left(\frac{1}{2}
	+i\nu\right)}{8 N_c}\log\left(\frac{1}{x}\right)
\left[-\psi\left(\delta-\frac{1}{2} - i\nu\right)
-\log\frac{\kappa^2}{M_h^2}\right]\right\}\,,
\]
where $\beta_0=(11 N_c-2 N_f)/3$, with $N_f$ the number of
active quarks,
$\bar{\alpha}_s = \alpha_s\left(\mu^2\right) N_c/\pi$,
with $\mu^2 = Q_0 M_h$, and $\chi_0(\frac{1}{2} + i\nu)$ is the LO eigenvalue of the BFKL
kernel. Here, $M_h$ plays the role of the hard scale which can be identified
with the photon virtuality, $\sqrt{Q^2}$.
In Eq.~\eqref{HentsUGD}, $\chi(\gamma)$
is the NLO eigenvalue of the BFKL kernel, collinearly improved and employing BLM method for the scale fixing given as in Section~2 of Ref.~\cite{Chachamis:2015ona}.

This UGD model is described through three parameters $Q_0$, $\delta$ and ${\cal C}$ which characterize a peculiar parametrization for the proton impact factor (see Ref.~\cite{Chachamis:2015ona} for further details).
We adopted here the so called {\em kinematically improved} values (see Section~III A of Ref.~\cite{Bolognino:2018} for further details).

\subsubsection{Golec-Biernat--W{\"u}sthoff' (GBW) UGD}
This UGD parametrization derives from the effective dipole cross section
$\sigma(x,r)$ for the scattering of a $q\bar{q}$ pair off a
nucleon~\cite{GolecBiernat:1998js}, through a Fourier transform and reads
\begin{equation}
	{\cal F}(x,\kappa^2)= \kappa^4 \sigma_0 \frac{R^2_0(x)}{8\pi}
	e^{\frac{-k^2 R^2_0(x)}{4}}\,.
\end{equation}
We refer to Ref.~\cite{GolecBiernat:1998js} for the details and discussion of the parameters of this model.
\subsubsection{Watt--Martin--Ryskin' (WMR) model}

%\begin{figure}[h]
%	\centering
	
%\end{figure}
The UGD introduced in~\cite{Watt:2003mx} reads
\[
{\cal F}(x,\kappa^2,\mu^2) = T_g(\kappa^2,\mu^2)\,\frac{\alpha_s(\kappa^2)}
{2\pi}\,\int_x^1\!dz\;\left[\sum_q P_{gq}(z)\,\frac{x}{z}q\left(\frac{x}{z},
\kappa^2\right) + \right.\nonumber
\]
\begin{equation}
	\label{WMR_UGD}
	\left. \hspace{6.5cm} P_{gg}(z)\,\frac{x}{z}g\left(\frac{x}{z},\kappa^2\right)\,\Theta\left(\frac{\mu}{\mu+\kappa}-z\right)\,\right]\,,
\end{equation}
where the term $T_g(\kappa^2,\mu^2)$, whose expression is given in Ref.~\cite{Watt:2003mx}, gives the probability of evolving from the scale $\kappa$ to the
scale $\mu$ without parton emission.
This UGD model depends on an extra-scale $\mu$, which we fixed at $Q$.

\begin{figure}[h]
	\centering
	\includegraphics[scale=0.35,clip]{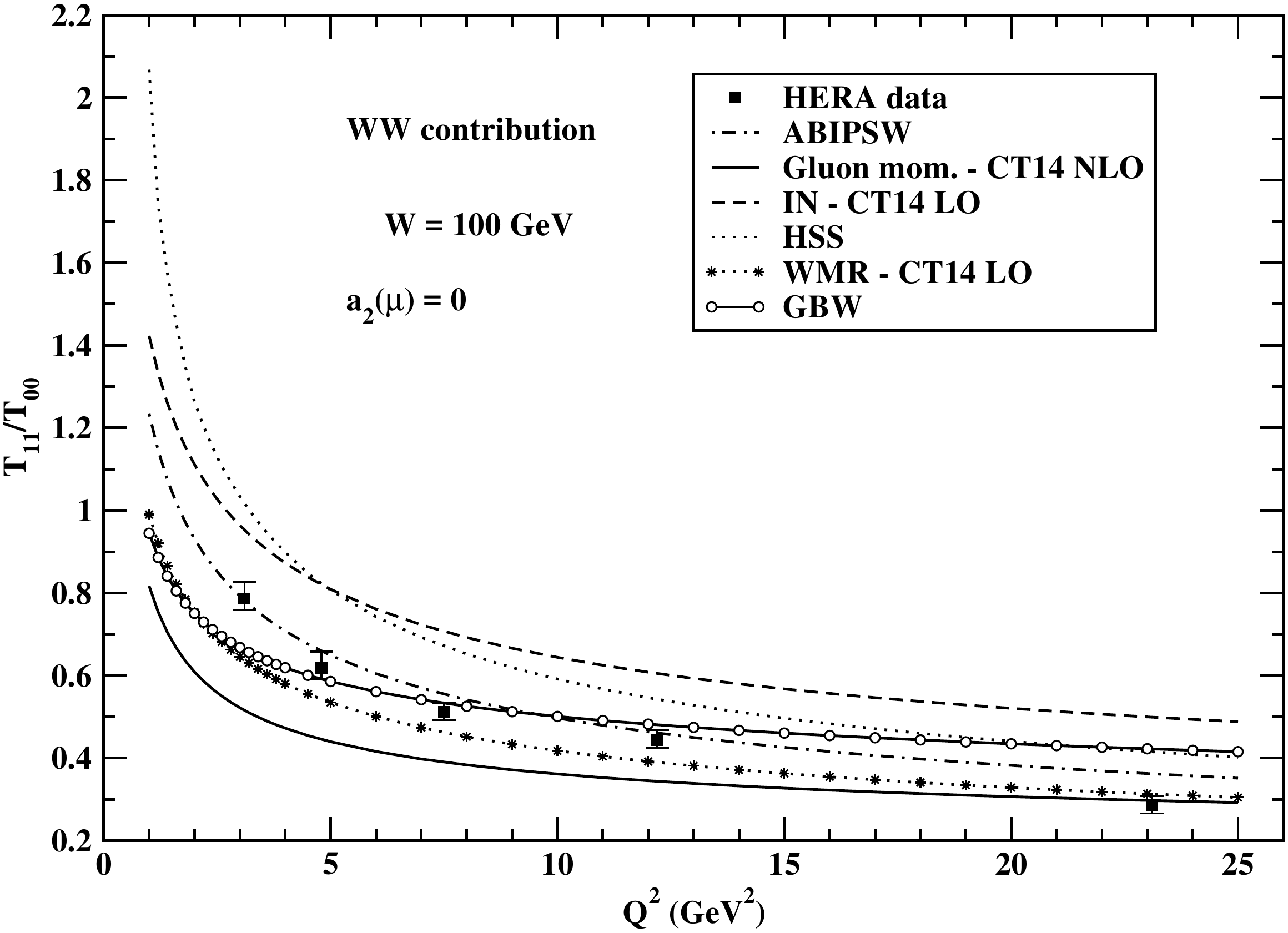}
	\caption{\emph{$Q^2$-dependence of the helicity-amplitude ratio $T_{11}/T_{00}$ for
			all the considered UGD models at $W = 100$ GeV. In the twist-2 DA we have
			put $a_2(\mu_0 = 1 \mbox{ GeV}) = 0$ and the $T_{11}$ amplitude has
			been calculated in the WW approximation.}}
	\label{fig:ratio_all}
\end{figure}
\begin{figure}[h]
	\centering
	\includegraphics[scale=0.35,clip]{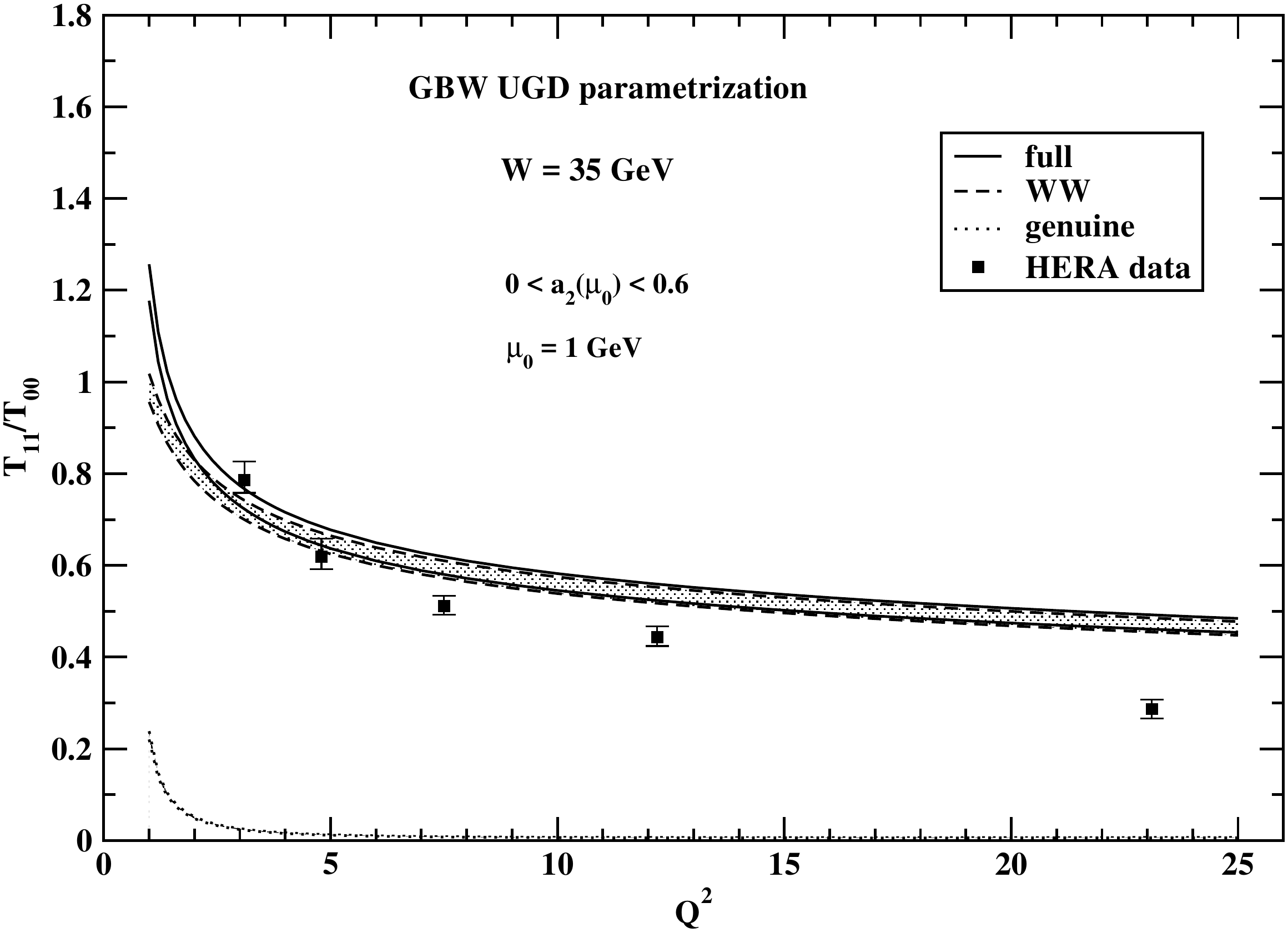}	
	\includegraphics[scale=0.35,clip]{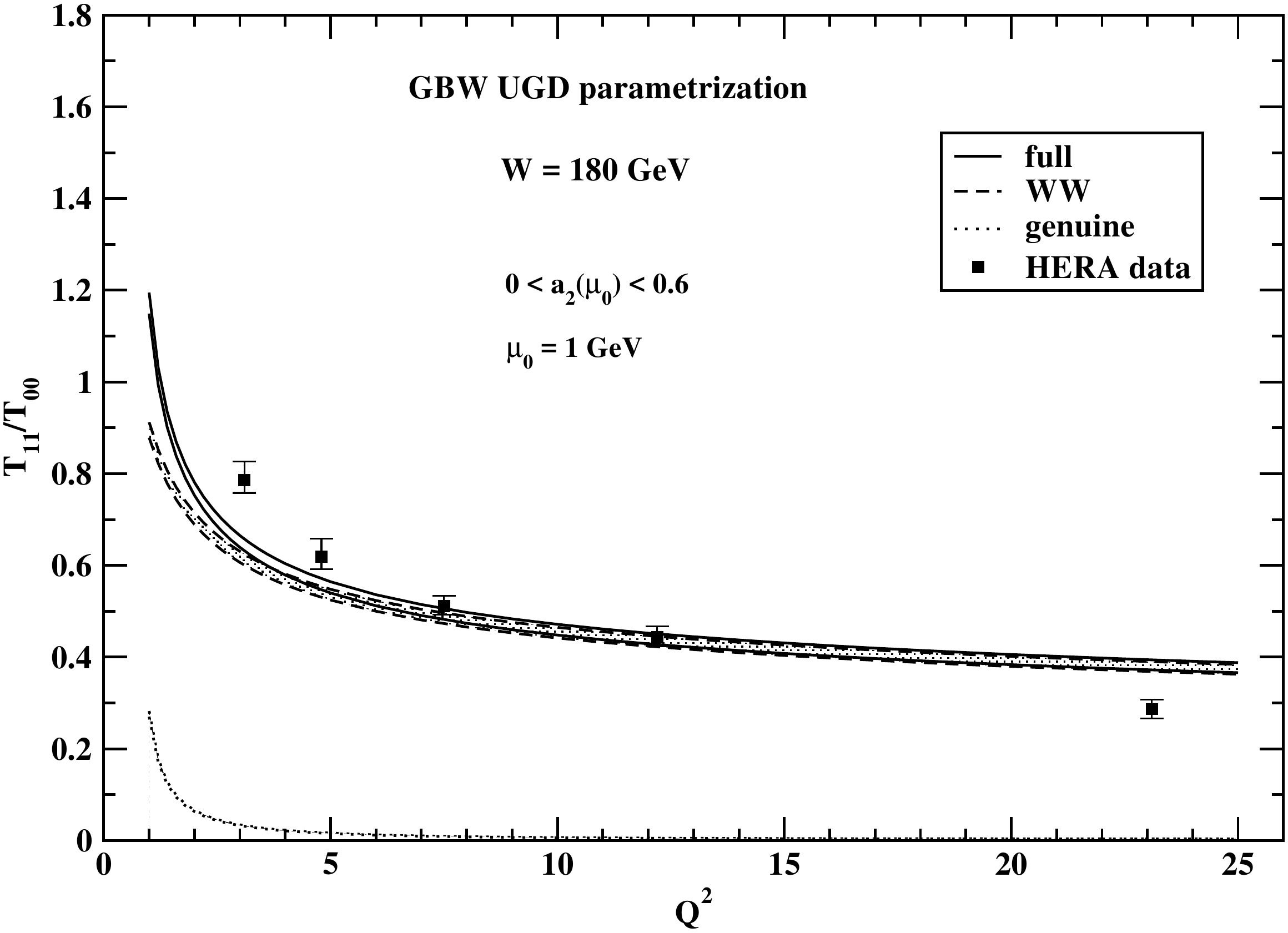}
	
	\caption{\emph{$Q^2$-dependence of the helicity-amplitude ratio $T_{11}/T_{00}$ for
			the GBW UGD model at $W = 35$ (left) and 180~GeV (right). The full, WW and
			genuine contributions are shown. The bands give the effect of
			varying $a_2(\mu_0 = 1 \mbox{ GeV})$ between 0. and 0.6.}}
	\label{fig:ratio_GBW_evolved}
\end{figure}
\begin{figure}[h]
	\centering	
	\includegraphics[scale=0.35,clip]{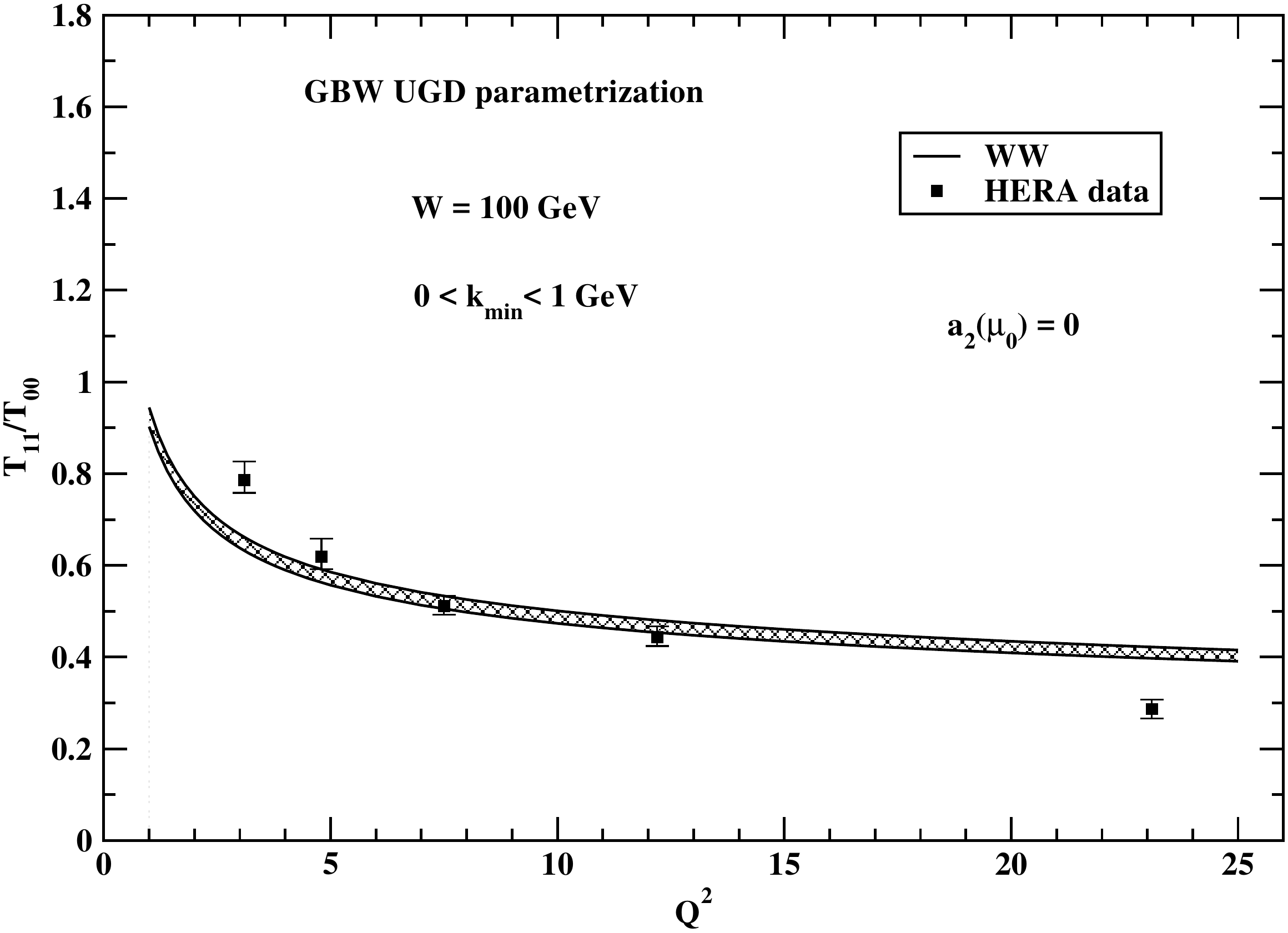}
	\caption{\emph{$Q^2$-dependence of the helicity-amplitude ratio $T_{11}/T_{00}$ for
			the GBW UGD model at $W = 100$~GeV. The band is the effect of a lower
			cutoff in the $\kappa$-integration, ranging from 0. to 1~GeV. In the
			twist-2 DA we have put $a_2(\mu_0 = 1 \mbox{ GeV}) = 0$ and the $T_{11}$
			amplitude has been calculated in the WW approximation.}}
	\label{fig:ratio_GBW_kmin}
\end{figure}
\section{Numerical analysis}
\label{analysis}

In this Section we present our results for the helicity-amplitude ratio
$T_{11}/T_{00}$, as obtained with the six UGD models presented above, and
compare them with HERA data.

In Fig.~\ref{fig:ratio_all} we compare the $Q^2$-dependence of $T_{11}/T_{00}$
for all six models at $W = 100$~GeV, together with the experimental result.

We used here the asymptotic twist-2 DA ($a_2(\mu^2)=0$) and the WW approximation
for twist-3 contributions. Theoretical results are spread over a large interval,
thus supporting our claim that the observable $T_{11}/T_{00}$ is potentially
able to strongly constrain the $\kappa$ dependence of the UGD. None of the
models is able to reproduce data over the entire $Q^2$ range; the
$x$-independent ABIPSW model and the GBW model seem to better catch the
intermediate-$Q^2$ behavior of data.

To gauge the impact of the approximation made in the DAs, we calculated
the $T_{11}/T_{00}$ ratio with the GBW model, at $W = 35$ and 180~GeV,
by varying $a_2(\mu_0=1 \ {\rm GeV})$ in the range 0. to 0.6 and properly
taking into account its evolution. Moreover, for the same UGD model, we
relaxed the WW approximation in $T_{11}$ and considered also the genuine
twist-3 contribution. All that is summarized in
Fig.~\ref{fig:ratio_GBW_evolved},  which indicates that our predictions for the helicity amplitude ratio are rather insensitive to the form of the meson DAs.

The stability of $T_{11}/T_{00}$ under the lower cut-off for $\kappa$, in the
range 0 $< \kappa_{\rm min} < 1$~GeV, has been investigated. This is a
fundamental test since, if passed, it underpins the main underlying
assumption of this work, namely that {\em both} the helicity amplitudes
considered here are dominated by the large-$\kappa$ region. In
Fig.~\ref{fig:ratio_GBW_kmin} we show the result of this test for the
GBW model at $W = 100$~GeV; similar plots can be obtained with the other
UGD models. There is a clear indication that the small-$\kappa$ region
gives only a marginal contribution.

%\begin{figure}[tb]
%	\centering
%	
%	\includegraphics[scale=0.60,clip]{x1d-3_UGDs.pdf}
%	
%	\includegraphics[scale=0.60,clip]{x1d-4_UGDs.pdf}
%	
%	\caption{$\kappa^2$-dependence of all UGD models for $x = 10^{-3}$ and $10^{-4}$.}
%	\label{fig:UGDs_vs_k2}
%\end{figure}

We refer to Section 3.1 of Ref.~\cite{Bolognino:2018} for details on the numerical implementation and on the systematic uncertainty estimation.
\section{Conclusions}
\label{discussion}

In this paper we have proposed the helicity amplitudes for the electroproduction
of vector mesons at HERA (and in possible future electron-proton colliders)
as a nontrivial testfield for models of the unintegrated gluon distribution
in the proton.

We have given some theoretical arguments, supported by a detailed numerical
analysis, that both the transverse and the longitudinal case are dominated
by the kinematic region where small-size color dipoles interact with the proton.
Moreover, we have shown that some of the most popular models for the
unintegrated gluon distribution in the literature give very sparse predictions
for the ratio of the transverse to longitudinal production amplitude.

\end{document}